\documentclass[10pt,letterpaper]{article}
\usepackage{gensymb} 

\def\XXint#1#2#3{{\setbox0=\hbox{$#1{#2#3}{\int}$}
     \vcenter{\hbox{$#2#3$}}\kern-.5\wd0}}

\def\a{\alpha}

\def\a0{\alpha_0}
\def\a{\alpha}

\def\r0{\rho_{0}}

\def\be{\begin{equation}}
\def\ee{\end{equation}}
\def\beq{\begin{equation}}
\def\eeq{\end{equation}}

\def\a0{\alpha_0}
\def\a{\alpha}

\def\r0{\rho_{0}}

\def\be{\begin{equation}}
\def\ee{\end{equation}}
\def\beq{\begin{equation}}
\def\eeq{\end{equation}}

\def\a{\alpha}

\def\({\left(}
\def\){\right)}

\usepackage[top=0.85in,left=2.75in,footskip=0.75in]{geometry}
\usepackage{amsmath,amssymb,changepage}
\usepackage[utf8x]{inputenc}
\usepackage{textcomp,marvosym}
\usepackage{cite}
\usepackage{nameref,hyperref}
\usepackage{microtype}
\DisableLigatures[f]{encoding = *, family = * }
\usepackage[table]{xcolor}
\usepackage{array}
\newcolumntype{+}{!{\vrule width 2pt}}
\newlength\savedwidth

\raggedright
\usepackage[aboveskip=1pt,labelfont=bf,labelsep=period,justification=raggedright,singlelinecheck=off]{caption}

\makeatletter
\renewcommand{\@biblabel}[1]{\quad#1.}
\makeatother

\usepackage{lastpage,fancyhdr,graphicx}
\usepackage{epstopdf}
\fancyheadoffset[L]{2.25in}
\fancyfootoffset[L]{2.25in}
\newcommand{\co}{\color{black}}

\graphicspath{{fig/}}
\begin{document}

\vspace*{0.2in}{\Large\textbf\newline{Optimizing running a  race on a curved track}}
\newline
\\
Amandine Aftalion\textsuperscript{1,*},
Pierre Martinon\textsuperscript{2},
\\
\bigskip
\textbf{1} Ecole des Hautes Etudes en Sciences Sociales, PSL Research University, CNRS UMR 8557,
Centre d’Analyse et de Math\'ematique Sociales, 54 Boulevard Raspail, Paris, France, amandine.aftalion@ehess.fr
\\
\textbf{2} Inria Paris and LJLL Sorbonne Universit\'e, Paris, France.
\\
\bigskip
* corresponding author

\section*{Abstract}
In order to determine the optimal strategy to run a race on a curved track according to the lane number,
we introduce a model based on differential equations for the velocity, the propulsive force and the anaerobic energy
which takes into account the centrifugal force.
This allows us to analyze numerically the different strategies according to the {\co types of track
since different designs of tracks lead to straights of different lengths.}
In particular, we find that the tracks with shorter {\co straights} lead to better performances,
while the double bend track with the longest {\co straight leads} to the worst performances and the biggest difference between lanes.
Then for a race with two runners, we introduce a psychological {\co interaction: there is an attraction to follow someone just ahead, but after being overtaken, there is a delay  before any benefit from this interaction occurs.
 We provide numerical simulations in different cases. Overall, the
 results agree with   the IAAF rules for lane draws in competition, where the highest ranked athletes get the center lanes, the next ones the outside lanes, while the lowest ranked athletes get the inside lanes.}

\section*{Introduction}
{\co In athletics,  inside  lanes are considered a disadvantage due to curvature, while in outside lanes, there is no one to chase.}
The aim of this paper is to understand from a physical and mathematical point of view the effect of the curved part of a track and
of the lane number on the running performance {\co both for a single runner and for a two-runner race}.

To our knowledge, no optimal control problem including these effects has been studied.
There is a huge literature on the way of running on a curved track, see for instance \cite{alex,greene,OTKH,ryan2003,UW,Quinn09,mur,AL}.
Nevertheless, it is never coupled with the psychological effect to have a neighbor on the next lane, which is mentioned as important.
Furthermore, though the IAAF regulations {\co\cite{iaaf}} do not impose a fixed shape of track, but allow the {\co straights } to vary between $80m$ and $100m$,
 we are not aware of any study discussing the effect of the {\co the lane and the track coupled with the psychological effect.}

In this paper, we will build on a model introduced by Keller \cite{keller1974optimal} and extended by \cite{aftalion_bonnans,aft},
to investigate how the shape of the track and the centrifugal force change the optimal strategy in a race: this leads to longer race
times for higher curvatures, and  therefore favors the {\co outer} lanes.
Estimating the performance of champions based on the modeling of Keller \cite{keller1974optimal}
has been developed  by various authors \cite{behncke1993mathematical,mathis1989effect,quinn2004effects,Quinn09,woodside1991optimal},
but never taking into account so many parameters as in this paper.
We  will also introduce a model taking into account the psychological effect between two runners.
 This is made up of two effects: on the one hand, the attraction by a runner close ahead, {\co and on the other hand, the  delay before any benefit  from the interaction occurs again after being overtaken}.
 This delay model is inspired by a paper on walking \cite{app}.
Let us point out that the mathematical problem encompassing delay in the equations is quite involved.  {\co We model the attraction by a runner close ahead as a  decreased friction, since the focus on chasing someone ahead improves the runner's economy.}
Due to the staggered start positions in the curved part of the track, this "rabbit effect" is less favorable on the {\co outer} lanes.

After introducing the model, we perform simulations using the optimal control toolbox \textsc{Bocop} \cite{Bocop}.
Since the IAAF regulations do not impose a single shape of track, we analyze the effect of the shape of the track on the optimal velocity profile, as well as the influence of the various parameters of the runner for a single runner. Then we perform simulations for two runners and
 our results show that the combination of the centrifugal force and the two runners interaction brings a numerical justification to the fact that the central {\co lanes} are the most
favorable to win a race.

\section*{Race model}

\subsection*{Model for a single runner race}
\label{model_1_runner}

\textbf{Single runner on a straight track.}
When a runner is running on a {\co straight}, as used by Keller \cite{keller1974optimal}, according to Newton's second law, the acceleration is equal to the sum of forces.
We can list two forces, the propulsive force $f(t)$ in the direction of motion, and the friction force, that we assume to be linear in velocity.
This leads to the first equation of motion for the velocity $v(t)$  written by unit of mass:
\beq\label{newton} \dot{v}(t) = f (t) - \frac{v(t)}{\tau}\eeq
where $\tau$ is the friction coefficient.
This coefficient {\co  models the friction due to various effects such as} joints friction, the runner's economy and the elasticity of the track. {\co Other friction effects can be encompassed in the equation such as wind or slopes, that is going upwards or downwards, or banking.}
Because the runner has a limited capacity, the propulsive force is bounded from above by a constant $f_{M}$, that is
\beq\label{fmax}0\leq f(t)\leq f_M.\eeq Typical values for $f_M$ range from 13 for a sprint at the world level \cite{morinaccel} to 5 for a marathon \cite{hoog}.

The power developed by the propulsive force is $f(t) \cdot v(t)$, which is to be taken into account in the energy balance.
This energy balance leads to the definition of the oxygen uptake $\sigma$ introduced in \cite{aftalion_bonnans}, which depends on the anaerobic energy $e(t)$.
Note that at the beginning of the race, {\co $e(0)=e_0$}, the available energy at initial time, and $\sigma$ depends on the accumulated oxygen deficit at time $t$, that is {\co $e_0-e(t)$.}
The function $\sigma$ depends on the length of the race \cite{hanoneffects}: for short races (up to 400m), $\sigma$ is a linear increasing function of {\co $e_0-e$}.
When the race gets longer, $\sigma$ reaches its maximal value $\sigma_{{max}}$ in the central part of the race, but is increasing at the beginning of the race,
and decreasing at the end.
Note that $\sigma$ is the energetic equivalent of $\dot{V O2}$, the volume of oxygen used by a unit of time and $\sigma_{max}$ is related to $\dot{V O2}_{max}$.
For the shorter races considered in this paper, we assume a linear function $\sigma$ and note $\sigma_f$ the final value, thus
\beq \label{sigma} \sigma (e)=\sigma_f {\co \frac {e_0-e }{e_0}}.\eeq
This leads to the energy model
\beq\label{energy} \dot{e}(t) = \sigma(e(t)) - f(t)v(t),\quad e(t)\geq 0,\quad {\co e(0)=e_0}.\eeq

A champion-level runner has a $\dot{VO2}_{max}$ about $75 ml/mn/kg$.
Since one liter of oxygen produces an energy about $21.1kJ$ via aerobic cellular mechanisms, this provides an estimate of the available energy per $kg$ per second $\sigma_{max} = 75/60*21.1 \approx 26 {\co m^2s^{-3}}$.
Furthermore, on a 200m race, $\dot{VO2}$ and $\sigma$ reach only about 75\% of their maximal values {\co\cite{hanoneffects,SG}}, so we set $\sigma_f = 20$.
We point out that this term is of lower order than $\dot{e}$: {\co we will see below in our simulations that the anaerobic part is roughly $87\%$ of the total energy, which is perfectly consistent with \cite{gastin}.}\\

For a fixed value of the final distance $d$, the optimal strategy to run the race is obtained by solving the control problem
(\ref{newton})-(\ref{fmax})-(\ref{sigma})-(\ref{energy}) under the constraint:
$$\hbox{minimize }T, \hbox{ such that } \int_0^T v(t) \ dt= d.$$
This problem has been studied in \cite{aftalion_bonnans,aft,aftaction}.
The parameters are matched to reproduce champions' races.
For a race less than 400m, when the function $\sigma(e)$ is decreasing, the velocity is increasing and then decreasing.
Indeed, the runner never has enough energy to maintain his maximal force for the whole duration of the race.
Therefore, the optimal strategy is to start at maximal force, and then the force decreases, and so does the velocity.

Since in the optimal control problem, it is usually the distance which is prescribed, in this paper,
we choose to take the distance $s$ instead of the time $t$ as variable.
We define $y(s)$ to be the time required to run the distance $s$ so that, if $x(t)$ is the distance run in time $t$, we have
\beq \dot{y}(s)=\frac 1 {v(t)} \hbox{ since }x(y(s))=s.\eeq
We call $f(s)$ the propulsive force needed at distance $s$ and $e(s)$ the energy.
This allows us to derive the equations for $y(s)$, $f(s)$, $e(s)$, from (\ref{newton})-(\ref{energy}), which are
\begin{eqnarray}
\label{eq2rr} \ddot{y} (s)= -f(s) \dot{y}^3(s)+\frac 1 \tau \dot{y}^2(s), & y(0) = 0,\hbox{ and } \dot{y}(0)=1/v^0, \\
\label{eq2r} \dot{e} (s)= \sigma (e(s))\dot{y}(s) - f(s),  & e(s) \geq 0,  \hbox{ and } e(0) = {\co e_0.}
\end{eqnarray}
This formulation requires an initial velocity $v^0$ which is not zero, but given the effect of the starting blocks where our dynamical model
is not correct, assuming an initial velocity of 3 or $4 m/s$ is quite consistent with the effect of the beginning of the race, 10m from {\co the start} \cite{sam15}.

The  constraint on the force is
\beq 0\leq f(s)\leq f_M\quad \hbox{ for }  0\leq s\leq d. \label{posenP}\eeq
The optimal control problem as such would lead to variations of the force which are too strong.
In order to take into account the impossibility for the runner to vary his propulsive force instantaneously, we instead take $df/ds$ as a bounded control.
We seek the optimal race strategy to minimize $T=y(d)$.\\

\textbf{Centrifugal force on a curved track.}
For races of $200m$ or more, the track is not  {\co straight but  includes one or more bends.
While on a bend,} the runner has to move against the centrifugal force, which, by unit of mass,
is $f_c = v^2/R$ where $v$ is the velocity of the runner and $R$ the curvature radius.
In order to produce a mathematical model for the dynamics in the curved part, we have to take into account the centrifugal force in Newton's law of motion
and project this equation on the 3 directions of motion.\\

Even on {\co straights}, there is an equation to be written in the $z$ direction: the reaction of the ground, $N$, is equal to the weight.
By the principle of action/reaction, the reaction of the ground is equal to the runner's propulsive force  in the $z$ direction.
Note that the runner does not have his feet on the ground all the time in the stride: he rather pushes (propulsive force) only for some time in a stride \cite{morinaccel}.
Some remarks in \cite{physrun} can be found related to this issue.
We point out that there is an interesting explanation of the effect of arms to counterbalance the torque, and that since there are two legs,
the reaction on each leg is not exactly the same \cite{alex}.
In this paper, we {\co do not include these effects as we believe them to be of minor importance.}
The specificity of our work is that although we consider a mean force and mean velocity in a stride, our model  allows us to
compute an instantaneous force and speed along the race.\\

\begin{figure}[ht!]
\begin{center}
 \includegraphics[width=0.8\textwidth]{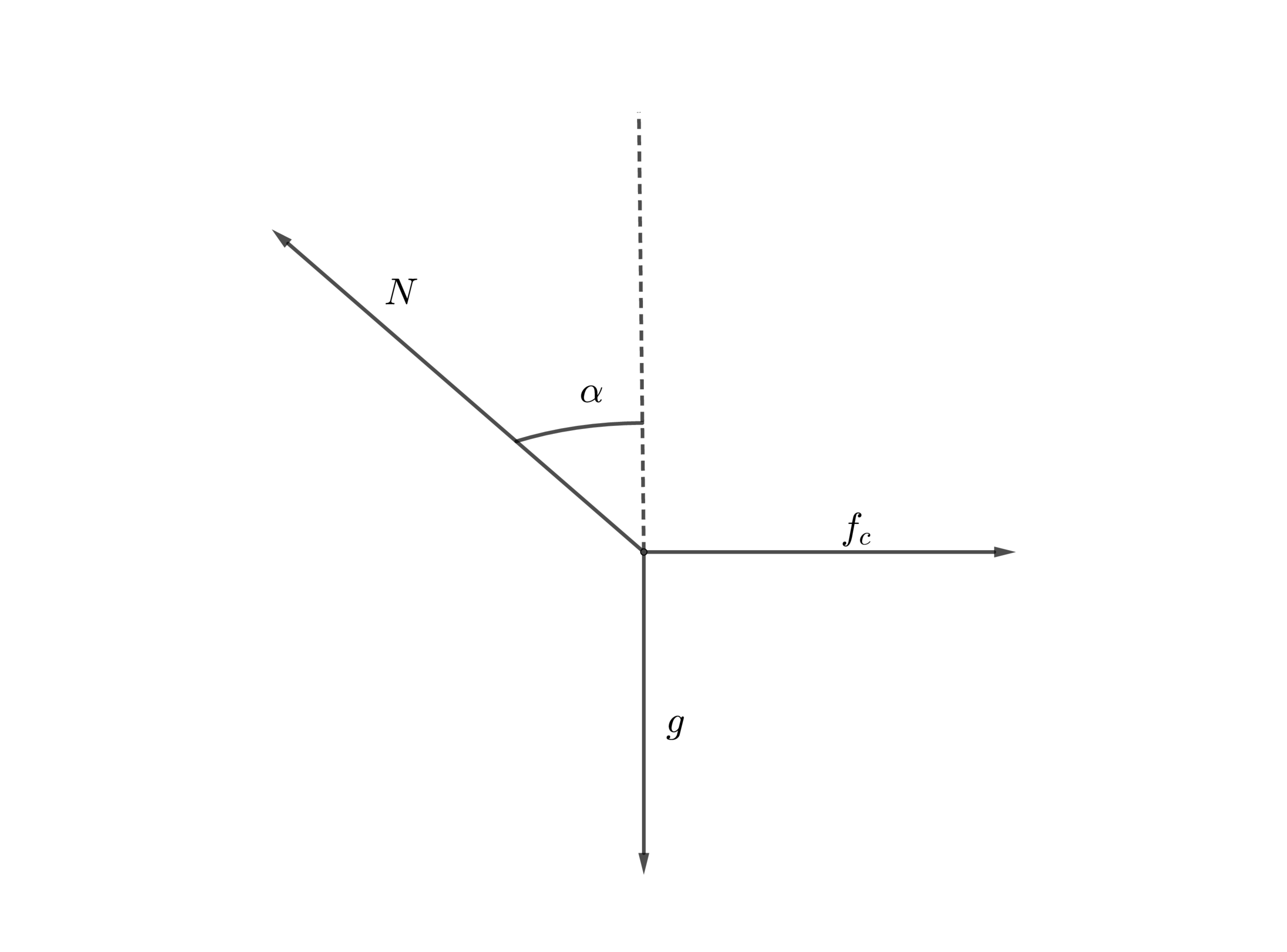}
\end{center}
\caption{\label{figforces} Illustration of the forces on the runner}
\end{figure}

On a curve, the runner {\co makes} an angle $\alpha$ with the vertical axis to balance the centrifugal force.
The runner is subject to gravity $g$, to the reaction of the ground $N$ along the angle $\alpha$, and to the centrifugal force $f_c=v^2/R$ (see Fig \ref{figforces}).
One has to consider the equations of motion
in the centrifugal direction and the $z$ direction, which lead to
\beq\label{equi}\frac{v^2}R =N \sin \alpha,\ g=N \cos \alpha\eeq
which provides the angle according to the velocity and the value of $N$:
\beq\label{equi2}\tan \alpha =\frac {v^2}{Rg},\ N ^2=g^2+\frac {v^4}{R^2}.\eeq
By the principle of action/reaction, the propulsive force in the transverse direction is the opposite of the reaction of the ground in the horizontal direction,
hence is equal to $N \sin \alpha$. Moreover, the propulsive force in the vertical direction is $N \cos \alpha$.
The total propulsive force $F$ is therefore such that $F^2=f^2 +N^2$ where we recall that $f$ is in the direction of movement.
From (\ref{equi2}), we find
\beq\label{centrif}F^2 = f^2+N^2= f^2+g^2+\frac {v^4}{R^2}.\eeq
Since $F$ has to be bounded and $g$ is constant, this leads to the new constraint
\beq\label{constcent}f^2+\frac {v^4}{R^2}\leq f_{M}^2.\eeq
We point out that eventually the effect of the centrifugal force is taken into account in the force constraint.
It cannot have an energy effect directly since the centrifugal force does not produce any work.\\

\begin{figure}[ht!]
 \begin{center}
  \includegraphics[width=0.9\textwidth]{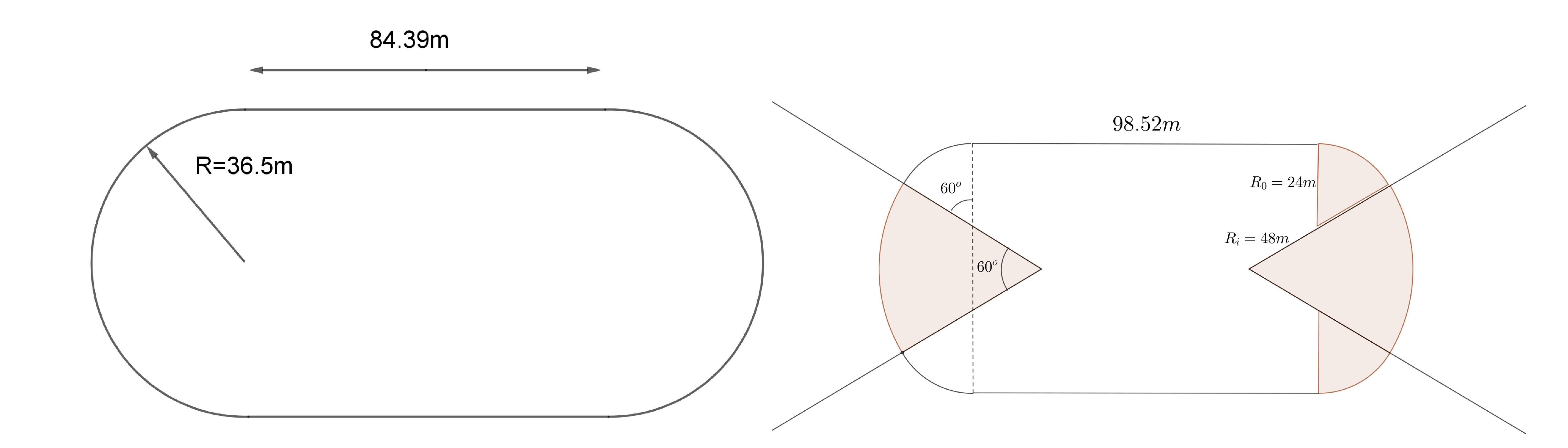}
  \vspace*{0.3cm}
  \caption{\label{figstadium} {\co Shape  for standard track (left) and and double bend 2 track (right)}.}
  \begin{tabular}{l|r|r|r}
  Track         & Straight  & Circle    & \\
  Standard      & $84.39m$  & $(36.50m,180\degree)$       &\\
  \hline
  Track         & Straight  & Circle 1  & Circle 2 \\
  Double Bend 1 & $79.996m$ & $(34.00m,2\times70\degree)$ & $(51.543m,40\degree)$\\
  Double Bend 2 & $98.52m$  & $(24.00m,2\times60\degree)$ & $(48.00m,60\degree)$\\
  \end{tabular}
 \end{center}
 \vspace*{-0.5cm}

\end{figure}

\textbf{Study of the track shape.}
It is important to know the exact shape of the track since it  influences the runner's optimal pacing strategy and performance.
However, there is no fixed regulation to build an athletic track.
Actually, as indicated in the IAAF manual \cite{iaaf}, the length of the straight part can vary between $80$ and $100m$,
while the curved part can be a half circle (`standard' track) or two different circular sections (`double bend' tracks).
We choose to study a standard track with an $84.39m$ straight part, and then two double bend tracks with straight parts of $79.996m$ and $98.52m$ respectively.
The shapes and dimensions of theses tracks are detailed in Fig \ref{figstadium}.
Note that for races longer than $100m$, runners start the race in the curved part.
The starting positions are therefore adjusted in order to have the same total distance for all lanes (`staggered start'). \\

Note that each runner is assumed to run at a distance of $30cm$ from the {\co inner} limit of the lane.
This is how the radius for the circular parts is set in order to obtain a $400m$ distance for lane 1. Then the width of each lane is $1.22m$.
This leads to different radii of curvature $R_k(s)$ depending on the lane $k$ and the distance $s$ run on the lane since {\co the start}.
On the straight part, $1/R_k(s)=0$. For more details on the value of $R_k(s)$ according to the track, we refer to \nameref{appendix:track_shapes}.\\

We want to point out that at the junction between the circular and straight parts, the runner will experience a discontinuity in the centrifugal force.
This force is $0$ on the straight part and can reach a value of the order of $2.5N$ per kilo on the circular part (since $v\sim 10 m/s$ and $R\sim 40m$),
which is about one quarter of the weight.\\

We will see on the numerical simulations that this may lead to an acceleration of the runner when reaching the straight part of the track.
One could think that it would be better to build a track where the curvature goes smoothly from 0 to the value of the matched circle so that the runner
experiences a continuous variation of his centrifugal force.
This type of curve, known as a clothoid, is used for instance for railways and roads.
The simulations in Section \nameref{simus_1_runner} indicate that the final time is actually larger on a clothoid,
because the smooth transition leads to a smaller radius for the circular part, therefore a larger centrifugal force.

One of the main results of our simulations is that the tracks with shorter {\co straights} lead to better performances 
(see \nameref{tracks}).\\

\textbf{Final model for a single runner race.}
The optimal problem is to minimize $T=y(d)$ with $y(s)$, $f(s)$, $e(s)$ solving (\ref{eq2rr})-(\ref{eq2r}), $\sigma$ being given by (\ref{sigma}),
with the bounded control
\beq\label{controldf}\left | \frac{df}{ds} \right |\leq 0.015\eeq
and the force constraint coming from (\ref{constcent})
\beq\label{boundfmcurve} f^2(s)+\frac 1{\dot{y}^4(s)R_k^2(s)}\leq f_M^2\eeq
where the curvature radius $R_k(s)$ is prescribed according to the lane $k$ and the track shape, see \nameref{appendix:track_shapes}.
We use the convention $R_k(s)=+\infty$ on a {\co straight}.\\

Finally, introducing a  state variable for the inverse of speed $z(s) = 1 / v(s)$, the optimal control problem for a single runner is
$$
(OCP)_1 \left \lbrace
\begin{array}{l}
\min\ y(d),\\
\dot y(s) = z(s) , \quad s \in [0,d], \quad y(0)=0,\\
\dot z(s) = z^2(s) / \tau - f(s) z^3(s) , \quad s \in [0,d],  \quad z(0)=\dot{y}(0)=1/v^0,\\
\dot e(s) = \sigma(e(s)) z(s) - f(s), \quad s \in [0,d], \quad {\co e(0)=e_0},\\
\dot f(s) = u(s), \quad s \in [0,d],\\
|u(s)| \le 0.015, \quad s \in [0,d],\\
e(s)\geq 0, \quad s \in [0,d],\\
f^2(s)+\frac 1{z^4(s) R_k^2(s)}\leq f_M^2, \quad s \in [0,d].
\end{array}
\right .
$$

\subsection*{Model for a two-runner race}
\label{model_2_runner}
When two runners are involved, we label them with $i$, $i=1,2$ and define $y_i(s)$, $f_i(s)$, $e_i(s)$ respectively the time to reach the distance $s$,
the propulsive force at distance $s$ and the anaerobic energy left at distance $s$. We also label by $i$ the parameters of each runner: $\tau_i$ the friction coefficient,
$f_{M,i}$ the maximal force, {\co $e_{0,i}$} the initial energy, $v_i^0$ the initial velocity.
Finally, we call $T_i$ the final time to reach the distance $d$ that is $T_i=y_i(d)$.\\

\textbf{Objective function.}
We want to solve the race problem where both runners try to obtain their minimum time and win the race.
The issue is to define a good mathematical problem.
Minimizing $min(T_1,T_2)$ is not enough since it could lead to a situation where one of the runner stops optimizing his race once he knows he will lose.
Then, minimizing the sum of the times $T_1+T_2$ could lead to some cooperative interaction where the faster runner would wait for the slower one to optimize the global time.
This is why we choose to minimize a combination of these two objectives, namely minimize
$$
min(T_1,T_2) + k_w (T_1+T_2)
$$
with $k_w$ being a small parameter {\co such that the second term does not modify the value of the leading order $min(T_1,T_2)$, but yet does not let $max( T_1,T_2) $, which is the time of the slower runner,  be too big. In our simulations, values of $k_w$ ranging from $10^{-3}$ to $10^{-4}$ provide this kind of behaviour. } \\
We point out that some authors \cite{appaft,hil} have tried to settle a stochastic description in the framework of game theory but they are not able to handle as many parameters as this model.
Also in a short race, we do not believe that there is time to think and adapt one's strategy on the course of the race.\\

\textbf{Psychological interaction.}
\label{interaction}
When two runners race against each other, we introduce an interaction term which mollifies the friction term of each runner $\dot{y}_i^2/\tau_i$.
This term is equal to 1 in case of no interaction, and is lower than $1$ in case of a beneficial interaction.
It models the psychological benefit that comes from chasing someone just ahead.
Note that this interaction is not an aerodynamic effect ('drafting') as in bicycle or car racing, because the velocity is too small.
Cognitive effects are known to reduce perceived exertion: {\co  shielding has a psychological basis for runners and the focus on chasing produces better running economy \cite{crews,stones,pugh}}.
 This psychological effect is indeed acknowledged by runners (sometimes called "rabbit effect") and can allegedly have an effect as high as 1 second per $400m$ lap {\co\cite{pugh}}.\\

The differential equations for $y_1$ and $y_2$ are therefore
\begin{eqnarray}
\ddot{y}_1 (s)= -f_1(s) \dot{y}_1^3(s)+\frac 1 {\tau_1} \dot{y}_1^2(s)(1 -  F(y_1(s),y_2(s))),\\
\ddot{y}_2 (s)= -f_2(s) \dot{y}_2^3(s)+\frac 1 {\tau_2} \dot{y}_2^2(s)(1 -  F(y_2(s),y_1(s)))
\label{eq2int}
\end{eqnarray}
where $F(y_1,y_2)$ is to be determined  as a function of $r(s)$ which is the distance between the two runners.
The detailed expression of $r(s)$ is presented in \nameref{appendix:distance_gap}.\\

\begin{figure}[ht!]
\begin{center}
 \includegraphics[width=0.5\textwidth]{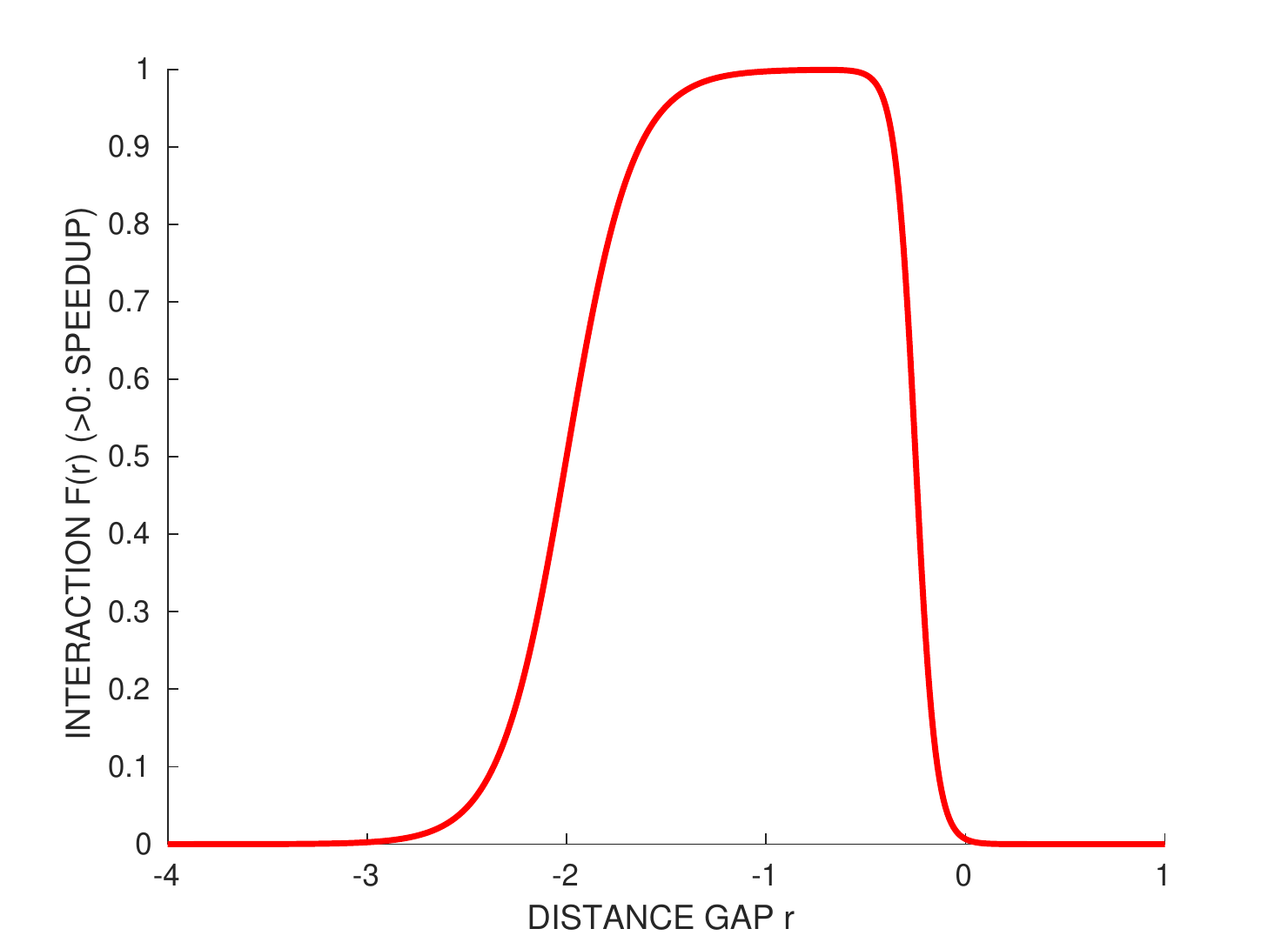}
\end{center}
\caption{\label{figinteraction} Illustration of the interaction between two runners (for $\gamma=1$)}
\end{figure}

\textbf{Basic interaction.}
We choose the  function $F$ of $r$ to be equal to 0.04 when $r$ is roughly between 0 and $-2.5m$ {\co and 0 outside this interval, which corresponds to the distance for which an effect can be felt. A lot of possible functions can match this goal. We choose for instance
} the interaction function illustrated in Fig \ref{figinteraction}
$$
F(r) = \gamma H(r+a_1,b_1,\epsilon) H(-r+a_2,b_2,\epsilon)
$$
where $\gamma=0.04$, $H$ a smoothed Heaviside function defined by
\beq\label{H}
H(r,k,\epsilon) = (1 + e^{-2k(r+\epsilon)} )^{-1}
\eeq
and with the values for the offsets and slopes $a_1=2$, $b_1=3$, $a_2=-0.25$,$ b_2=10$, and $\epsilon=10^{-6}$.\\

\textbf{Lateral attenuation.}
It seems reasonable to assume that the positive interaction only occurs when the two runners are close enough to each other.
Thus we introduce a limitation of the interaction based on the lane gap between the runners.
In practice, the interaction is multiplied by an attenuation function $A(k_1,k_2)$
\beq\label{interF}
F(r) = \gamma A(k_1,k_2) H(r+a_1,b_1,\epsilon) H(-r+a_2,b_2,\epsilon)\eeq
with $A$ defined by
$$
\begin{array}{lll}
A(k_1,k_2) = & 0                        & \quad if\ |k_1-k_2| \ge 4\\
             & 1 - \frac{|k_1-k_2|}{10} & \quad otherwise.
\end{array}
$$

\textbf{Inhibition and delay problem.}\\
A refinement of the interaction model is that the benefit should not hold just after being overtaken, since there is a delay in reacting.
In order to build a mathematical model for this, we use the behavior of following in pedestrian traffic introduced in \cite{app}.
The delay is meaningful in terms of human perception: perception of speed actually comes from successive perceptions of distance over time,
and this integration process introduces a delay, while the perception of distances can be considered instantaneous.\\

We therefore introduce an inhibition formula that suppresses the interaction for a short duration after being overtaken.
Since our model is formulated with distance as the independent variable, the delay is also expressed in terms of a distance frame $\eta$.
The basic idea is to multiply the interaction term $F$ by an characteristic function $I_\eta$ defined by\\
\beq\label{eqI}
\begin{array}{lll}
I_\eta(s) = & 0  & \quad {if\ an\ overtaking\ has\ occured\ on}\ [s-\eta,s]\\
            & 1  & \quad {otherwise.}
\end{array}
\eeq
For the numerical simulations, $I_\eta$ is smoothed using the Heaviside approximation $H$ defined by (\ref{H}).
The detection of an overtaking is performed by checking for sign changes of $r$ over $[s-\eta,s]$.\\

Note that this check relies on \textbf{past} values of the state variables used to compute $r$, thus leading to a \textbf{delay} optimal control problem.
Delay problems are a quite involved class of optimal control problems, and we refer the interested readers to \cite{delay14,delay16} for recent theory developments.
A classical way to solve delay problems is to reformulate them as non-delayed problems, see \cite{Guinn1976}, but the manipulation is rather cumbersome.
In our case, we take advantage of the feature from the toolbox \textsc{Bocop} to handle delays automatically in the fixed final time case
(which we have since we use distance instead of time as the independent variable).\\

\textbf{Final model for a 2-runner race.}We define $T_i=y_i(d)$ and $F$ from (\ref{interF}).
The optimal control problem becomes:
$$
(OCP)_2 \left \lbrace
\begin{array}{l}
\min (\min(T_1,T_2) + k_w (T_1+T_2)),\\

\dot y_i(s) = z_i(s) , \quad s \in [0,d], \quad y_i(0) = 0,\ z(0)=\dot{y}_i(0)=1/v^0_i,\quad i=1,2,\\
\dot e_i(s) = \sigma(e_i(s)) z_i(s) - f_i(s), \quad s \in [0,d], \quad e_i(0) = {\co e_{0,i}},\quad i=1,2,\\
\dot f_i(s) = u_i(s), \quad s \in [0,d], \quad i=1,2,\\
\dot z_1 (s) = - f_1(s) z_1^3(s) + \frac 1 {\tau_1} z_1^2(s) (1 - {I_\eta(s)} F(r(s))),  \quad s \in [0,d],\\
\dot z_2 (s) = - f_2(s) z_2^3(s) + \frac 1 {\tau_2} z_2^2(s) (1 - {I_\eta(s)} F(-r(s))), \quad s \in [0,d],\\
e_i(s)\geq 0, \quad s \in [0,d], \quad i=1,2,\\
|u_i(s)| \le 0.015, \quad s \in [0,d], \quad i=1,2,\\
f_i^2(s)+\frac 1{z_i^4(s) R_{k_i}^2(s)}\leq f_{M,i}^2, \quad s \in [0,d], \quad i=1,2.

\end{array}
\right .
$$

It is worth pointing out that this optimal control problem has several families of local solutions,
typically with a different number of overtakings. In the numerical simulations, we overcome this
difficulty by trying several initial points and picking the best solutions.
Using a global optimization method would of course solve this problem, however in our case the dimension
of the state variables is too high.

\section*{Numerical simulations for a single runner}
\label{simus_1_runner}
In the numerical simulations, we chose to study the $200m$ race.
For reference, in 2018, the world record for $200m$ is 19.19s (Usain Bolt, Berlin World Championships, 2009).
In all the following, we will simulate races with fictitious runners whose parameters (see Table\ref{tab:runners})
are chosen so that their race times are close to 20s. {\co We have chosen the maximal force $f_M$ to range between $6.5$ and $13$ which is the range in the literature \cite{OTKH,sam15,Quinn09,mur,AL}. Then $\tau$ is chosen so that the peak velocity which is close to $f_M \tau$ is roughly 11.1, which is the world's level. The initial energy $e_0$ is such that the ratio of the anaerobic contribution to the total contribution is $87\%$ for runner $A_1$, which is consistent with \cite{gastin,SG}, and the final time is around $20$ seconds. The initial velocity is taken to be 1/0.43 to
 take into account the departure in the starting blocks \cite{sam15}. The bound on the maximal variation of $f$ is taken of order $\sigma_f/\tau e_0$, which is close to what we expect as a singular control \cite{Tre}.}
 \begin{table}[ht!]
\begin{center}
\caption{\label{tab:runners}Athletes' parameters. $A_1$ is a good runner, $A_2$ has a very strong propulsive force and $A_3$ is a poor runner.}
\begin{tabular}{l|rrrrc}
Runner  & $\tau$  & $e_0$ & $f_M$  &  $1/v^0$ & $|df/ds|_{max}$ \\
\hline
$A_1$   & 1.18    & 1500    & 9.45 & 0.43 & 0.015\\
$A_2$   & 0.85    & 2160    & 13   & 0.43 & 0.015\\
$A_3$   & 1.7     & 1000    & 6.5  & 0.43 & 0.015
\end{tabular}
\end{center}
\end{table}

\subsection*{Single runner on a straight track}
We start with a simple straight $200m$ race to illustrate the effect of parameters $f_M$, $\tau$, and {\co $e_0$.}
We take as reference athlete $A_1$ of Table \ref{tab:runners}.
The corresponding speed and force profiles are shown with black lines in Fig \ref{figrunner4_std}.
The velocity increases to its peak value $v_m\sim f_M \tau$ and then decreases.
The runner does not have enough energy to run the whole duration of the race at maximal force.

The propulsive force starts at its maximal value $f_M$, then decreases at the constant rate $|df/ds|_{max}$.
The time at which the force begins to decrease depends on the values of $f_M$ and {\co $e_0$.}
Indeed, increasing {\co $e_0$} does not change the beginning of the race but allows to run longer at $f=f_M$.
On the other hand, increasing $f_M$ increases the peak velocity but does not change much the second part of the race.
Finally, increasing $\tau$ has a more uniform effect and increases the velocity for the whole duration of the race.

\subsection*{Single runner on a standard curved track}

\begin{figure}[ht!]
\begin{center}
\includegraphics[width=\textwidth]{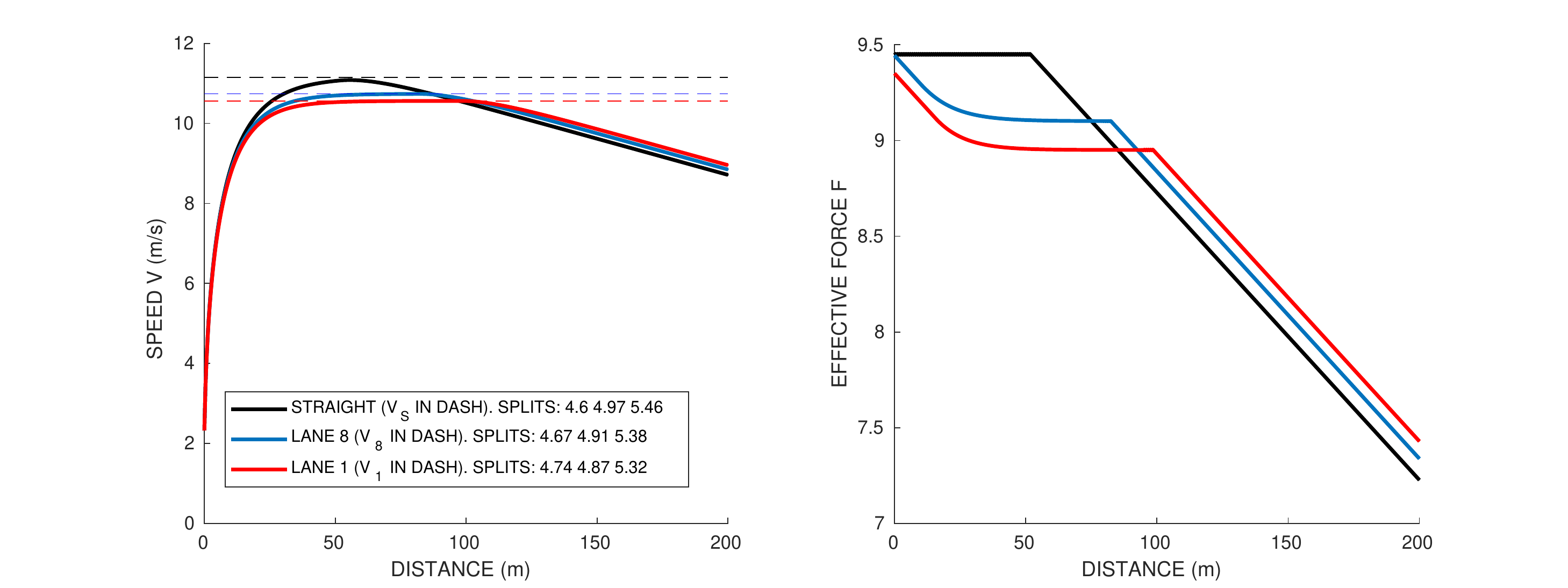}
\end{center} \vspace*{-0.3cm}
\caption{\label{figrunner4_std}Single runner $A_1$ on a standard track, lanes 1 and 8, and straight track. Force in $N/kg$ vs distance on the right graph.
Speed vs distance on the left graph, with the constant speeds given by eq. (\ref{vconst}) in dashed lines.
Time splits for $50-100m$, $100-150m$ and $150-200m$ are indicated.}
\end{figure}

We simulate the same runner on the so-called standard track, i.e. $115.61m$ half circle of radius $36.80m$ followed by a $84.39m$ {\co straight}.
Fig \ref{figrunner4_std} shows the race profiles obtained for the {\co inner and  outer} lanes (respectively 1 and 8), and the {\co straight} race.
The time splits for $50-100m$, $100-150m$ and $150-200m$ are indicated in the figure: we have chosen the parameters for $A_1$ so that
they match the order of magnitude of time splits for athletes in World Championships.
The velocity profiles of the curved track are quite different from the straight track:\\

i) the runner starts slower because of the curvature: even though he puts his maximal propulsive force at the start,
part of it is used to counterbalance the centrifugal force, resulting into a lower effective force and a lower velocity
$$f_M^2\geq f_{init}^2+\frac {(v^0)^4}{R_k^2}.$$\\

ii) in the middle part of the race, the maximal propulsive force is reached and we can derive from (\ref{newton}) and (\ref{constcent}) the relation between $f$ and $v$:
\beq v = f \tau \hbox{ and } f_M^2=f^2+\frac {f^4\tau^4}{R_k^2}\label{vkR}\eeq
with $R_k$ the curvature radius on lane $k$. We can compare this formula with our simulations: on the {\co straight} $v_s=f_M\tau$,
while from (\ref{vkR}), the velocity in the middle of the race on lane $k$ is
\beq v_s^2=v_k^2+\frac {v_k^4 \tau^2}{R_k^2}, \hbox{ that is } v_k^2=\frac{-R_k^2+R_k\sqrt{R_k^2+4v_s^2\tau^2}}{2\tau^2}\label{vconst}\eeq

The numerical simulations indicate an extremely good consistency with this expression: in this case equation (\ref{vconst}) yields
$v_s=11.15$, $v_8=10.74,$ and $v_1=10.56$, which are drawn as dashed lines in Fig \ref{figrunner4_std}.\\ 

iii) after the curved part, there is no more centrifugal force so that the runner can increase both his propulsive force and velocity.\\

iv) finally, at the end, the runner slows down again, because he does not have enough energy left to sustain his maximal force.\\

If we compare lane 1 and lane 8, on lane 1 the runner starts slower since the centrifugal force is stronger due to larger curvature.
On the other hand, he puts a slightly larger force in the second part of the race, having more energy left, yet he is slower overall.
Final times are: $20.43$ for the straight track, $20.46$ for lane 8, and $20.48$ for lane 1. {\co Let us point out that our simulations are consistent with the experiments in \cite{OTKH}, where runners are asked to run 60m on a straight path and on a curved path.
 The authors observe the existence of two groups, one "good" group  who manages to reach the same velocity in the curved path as in the straight path and the other "poor" group who is strongly affected by the curve.  Our parameters values of runner $A_1$ corresponds to a runner of the "good" group.}\\

Now we simulate several runners (see Table \ref{tab:runners} for parameters) in order to assess the influence of the maximal force $f_M$.

\textbf{Runner with large maximal force $f_M$.} We want to point out that due to the way the curvature is taken into account in the model,
see (\ref{constcent}), a runner with a greater maximal force $f_M$ will be less sensitive to the curvature of the track.
We illustrate this with the runner $A_2$ defined in Table \ref{tab:runners} whose $f_M=13$; final times are: $20.31s$ for straight track, $20.32s$ for lane 8 and 1.
In this extreme case, the runner is basically unaffected by the curvature of the track, that is the curves of velocity and force versus distance are almost the same for {\co straights}, line 1 and 8.\\

\begin{figure}[ht!]
\begin{center}
\includegraphics[width=\textwidth]{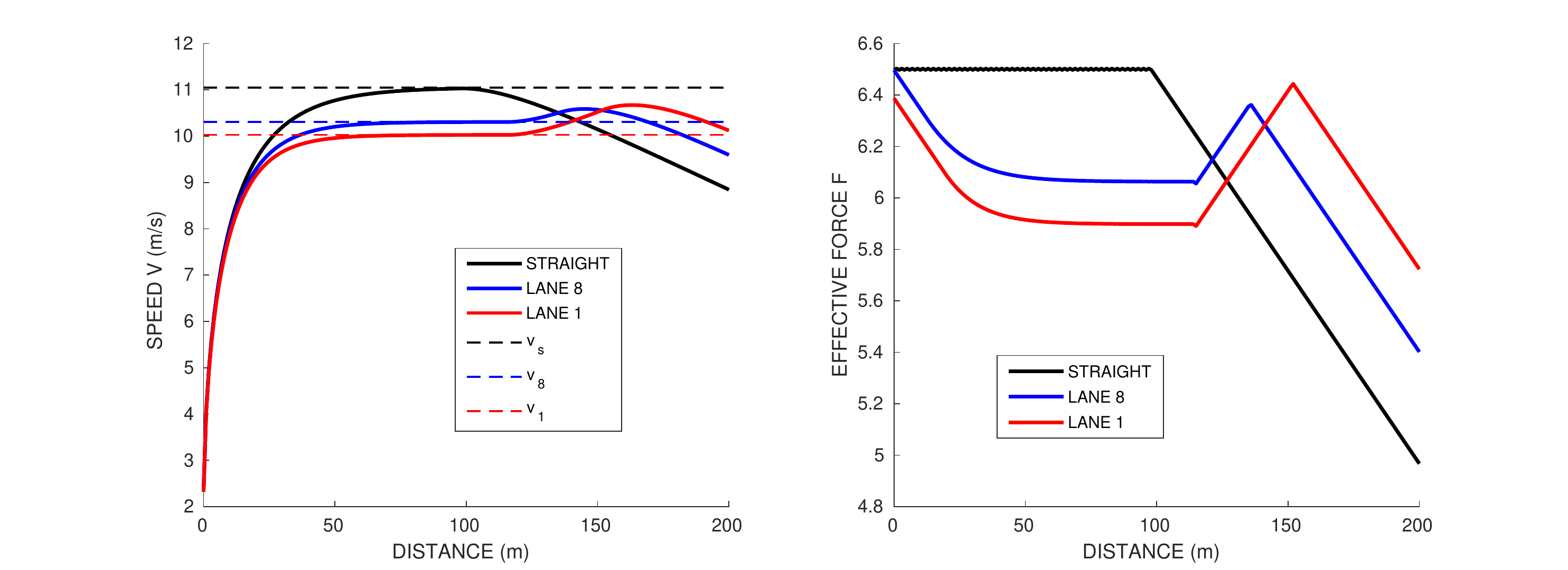}
\end{center} \vspace*{-0.3cm}
\caption{\label{figrunner1_reference}Single runner $A_3$ on a standard track, lanes 1 and 8, and straight track. Force in $N/kg$ vs distance on the right graph.
Speed vs distance on the left graph, with the constant speeds given by eq. (\ref{vconst}) in dashed lines.
 Force and velocity increase when the centrifugal force disappears.}
\end{figure}

\textbf{Runner with small maximal force $f_M$.}
With a low maximal force $f_M=6.5$, the runner $A_3$ of Table \ref{tab:runners} can increase his force and velocity when he reaches the straight part of the track, since he has not spent as much energy as the others at the beginning,
yet he is slower overall. This is illustrated in Fig \ref{figrunner1_reference}.
 Final times are: $20.32s$ for {\co straights}, $20.54s$ for lane 8, and $20.72s$ for lane 1. {\co  This runner $A_3$ corresponds to a runner of the "poor" group of \cite{OTKH}, with as much as $0.2s$ gap between extreme lanes. It is also consistent with performances for runners in \cite{Quinn09,mur,AL,morton}. }

{\co Let us point out that for these three runners, we have computed the ratio of the anaerobic energy versus the total energy and find: $87\%$ for $A_1$, $90\%$  for $A_2$ and $82\%$  for $A_3$, which is quite consistent with \cite{gastin}.}

\subsection*{Effect of different track shapes}
\label{tracks}
Now we study the effect of different track shapes defined in Fig \ref{figstadium}: standard with $84.39m$ straight (STD),
double bend 1 with $80m$ straight (DB1), double bend 2 with $100m$ straight (DB2),
and two modified standard tracks with smoothed curvature, including clothoid junctions of $10m$ (CL1) and $30m$ (CL2).
For the clothoid tracks, we choose a {\co straight} of $84.39m$ as the standard track. As explained in the Appendix, the length of the junctions provides the radius of the circle and the angle, which are respectively $33.32m$ and $164\degree$ for (CL1) and $29.95m$ and $118\degree$ for (CL2).

\begin{figure}[ht!]
\begin{center}
\includegraphics[width=0.9\textwidth]{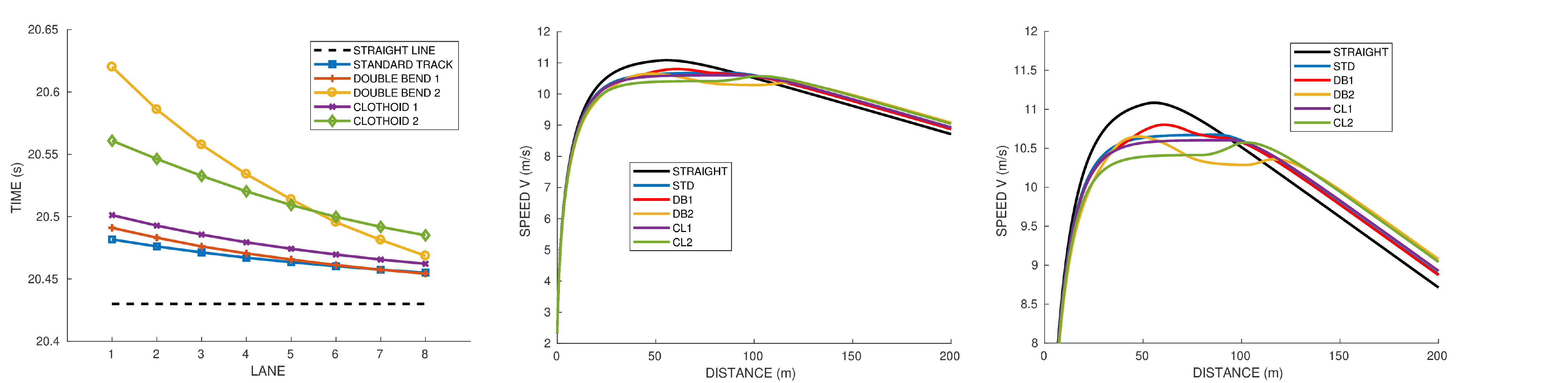}
\end{center} \vspace*{-0.3cm}
\caption{\label{figrunner4_tracks_times} Effect of the track shape on the race time  for Runner $A_1$ vs lane number (left graph). Speed profile for lane 5 (right graph) with zoom.
The tracks are standard (STD), double bend 1 (DB1) with short {\co straights}, double bend2 with long {\co straights} (DB2), and curves with short and long clothoid junctions (CL1 and CL2).}
\end{figure}

For each track shape, we simulate the race on the {\co inner  and outer} lanes (1 and 8).
The results are summarized in Table \ref{tab:tracks_times}, with the races for runner $A_1$ (on lane 5) shown in Fig \ref{figrunner4_tracks_times}.
Reference athlete $A_1$ has a difference of $0.17s$ between the best (DB1 track, lane 8) and worst (DB2, lane 1) case.
As mentioned previously, runner $A_2$ with a very high force $f_M=13$ is almost unaffected by the curvature, with times varying only between $20.32s$ and $20.36s$.
Yet, the DB2 track is still worse than the others.
Conversely, athlete $A_3$, with a lower force $f_M=6.5$, is more affected, with $1.01s$ between the best and worst cases.
The DB2 is his worst track and his best performance is on the standard track.

Our results show a time difference between {\co inner} and {\co outer} lanes ranging from $0.02s$ for the standard track to $0.15s$ for the worst double bend track.
This is consistent with \cite{Quinn09} who also finds the double bend track to be the worst, using a simplified model based on constant mean velocity and curvature.
 {\co Let us point out that in the next section, we will study a two-runner race where the effect of the lane becomes more pronounced: we find  a larger difference between the best and worst mean time per lane.}

Focusing on runner $A_1$ in Fig \ref{figrunner4_tracks_times}, we analyze more closely the effect of  the track shape and lane:\\
- DB1 is the quickest track for the {\co outside} lane, though it is very close to STD.\\
- The standard track has the  smallest difference between lanes.\\
- DB2 is the slowest track, from $0.01s$ on the {\co outside lane} to $0.14s$ on the {\co inner lane}.
When on the outer radius of curvature $24m$, the velocity significantly decreases.\\
- CL1 is quite close to DB1 and STD, though a little slower.  CL2 is slower than DB2 on the {\co outer} lanes, although not as bad in terms of difference between lanes.\\
It may seem surprising that the tracks with smoothed curvature do not perform better than the ones with a discontinuous curvature.
This comes from the fact that the clothoid junction actually results in a smaller radius for the circular part, and thus a greater curvature.
The longer the clothoid junction, the more pronounced the effect, and the slower the times.\\

To conclude the single runner races, it appears that the track with the shortest {\co straight} is the quickest track for strong athletes in {\co outer} lanes.
The standard track shape is the one with the best race times overall, and also the smallest time gap between the {\co inner} and {\co outer} lanes.
On the opposite, the double bend with the long $100m$ straight (DB2) yields the worst times overall, and the highest gap between the {\co inner} and {\co outer} lanes.
These conclusions seem consistent with runners' feelings though there is no study yet of what the ideal shape of track would be for a specific runner.

\begin{table}[ht!]
\begin{center}
\caption{\label{tab:tracks_times}Times for different runners and track shapes.}
{\scriptsize
\begin{tabular}{ll|rrrrr|r}
 runner    & shape:   & STD   & DB1   & DB2   & CL1   & CL2   & Straight\\
 \hline
 $A_1$     & lane 1   & 20.48 & 20.49 & 20.62 & 20.50 & 20.56 & 20.43\\
 $A_1$     & lane 8   & 20.46 & 20.45 & 20.47 & 20.46 & 20.49 & 20.43\\
 \hline
 $A_2$     & lane 1   & 20.32 & 20.33 & 20.36 & 20.33 & 20.34 & 20.31\\
 $A_2$     & lane 8   & 20.32 & 20.32 & 20.32 & 20.32 & 20.33 & 20.31\\
 \hline
 $A_3$     & lane 1   & 20.72 & 20.80 & 21.55 & 20.80 & 21.03 & 20.32\\
 $A_3$     & lane 8   & 20.54 & 20.55 & 20.66 & 20.57 & 20.63 & 20.32\\
\end{tabular}
}
\end{center}
\end{table}

\section*{Numerical simulations for two runners}
\label{simus_2_runner}

We move to the simulations for two-runner races, combining the interaction effect with the curvature effect previously studied for the single runner case.
Firstly, we study races with two  runners competing in adjacent lanes, to see the effect of the  interaction.
Then we compute the mean times corresponding to all possible races of a runner versus himself and find that the best lanes are indeed the center ones.


\subsection*{Races on different lanes and illustration of the interaction effect}

We perform simulations for the optimal control problem (OCP2) for two runners, combining the interaction effect with the centrifugal force.
We first set $A_1$ to be the runner on each lane 1 and 2.

\begin{figure}[ht!]
 \begin{center}
  \includegraphics[width=0.9\textwidth]{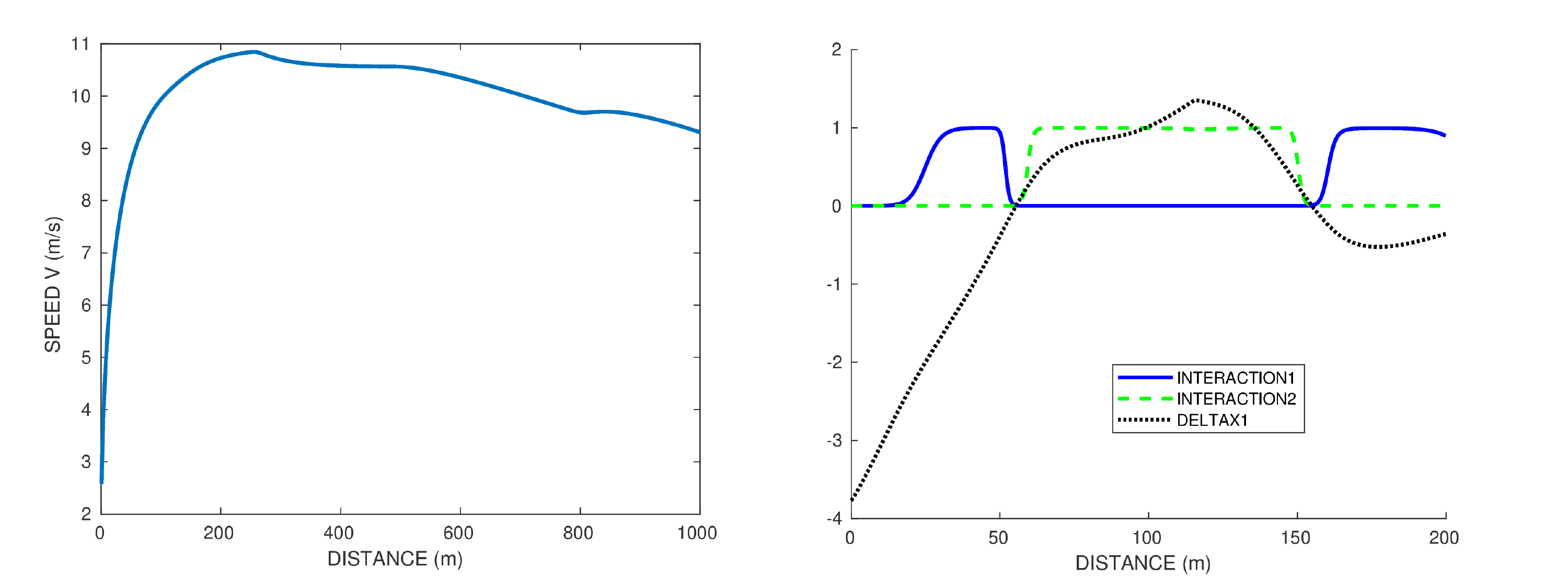}
 \end{center}\caption{\label{figrace12} Race $A_1$ vs himself at lanes 1-2, with interaction $\gamma=0.04$ and a frame $\eta=0$ that is no inhibition.
Left graph: speed profile and time splits of the runner at lane 1.
Right graph: distance gap and interaction term for both runners. The sign change of the distance gap corresponds to the overtaking. Lane 2 wins {by $0.04s$}.}\end{figure}

We recall that if $A_1$ runs alone, his time on lane 1 is $20.485s$ and on lane 2 $20.480s$, so of course because of the centrifugal effect, lane 2 is quicker.
Due to staggered starts, as soon as we set the interaction, the runner on lane 1 benefits from the interaction at the beginning of the race.
First, we set the interaction term $\gamma=0.04$ but with no inhibition $\eta=0$. The results are illustrated in Fig \ref{figrace12}, with the velocity profile in lane 1 on the left and the interaction for each runner and relative distance on the right.
When the relative distance is negative, the runner in lane 1 is behind. So in this case, the runner in lane 2 wins the race and they overtake each other twice:
lane 1 starts behind because of the staggered starts, benefits from interaction and overtakes at $50m$; then lane 2 benefits from interaction right away and is able to overtake again at $150m$.
Then they are on the {\co straight}, very close to each other, lane 1 benefits from interaction and is ready to overtake again but loses in the end by $0.04s$.

\begin{figure}[ht!]
 \begin{center}
  \includegraphics[width=0.9\textwidth]{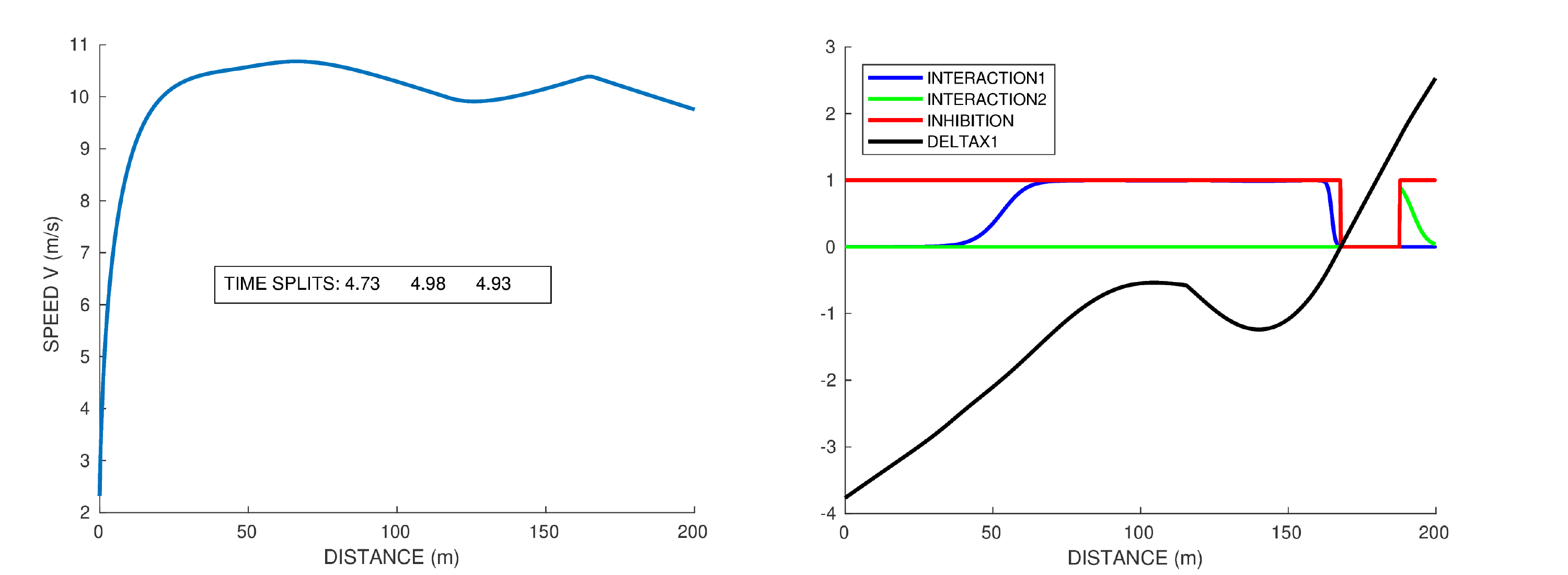}
 \end{center}
\caption{\label{figinhib} Race $A_1$ vs himself at lanes 1-2, with interaction $\gamma=0.04$ and a frame $\eta=20m$ for the inhibition.
Left graph: speed profile and time splits of the runner at lane 1.
Right graph: distance gap and interaction term for both runners. The sign change of the distance gap corresponds to the overtaking. Lane 1 wins {by $0.27s$}.}
\end{figure}

\begin{figure}[ht!]
 \begin{center}
  \includegraphics[width=0.9\textwidth]{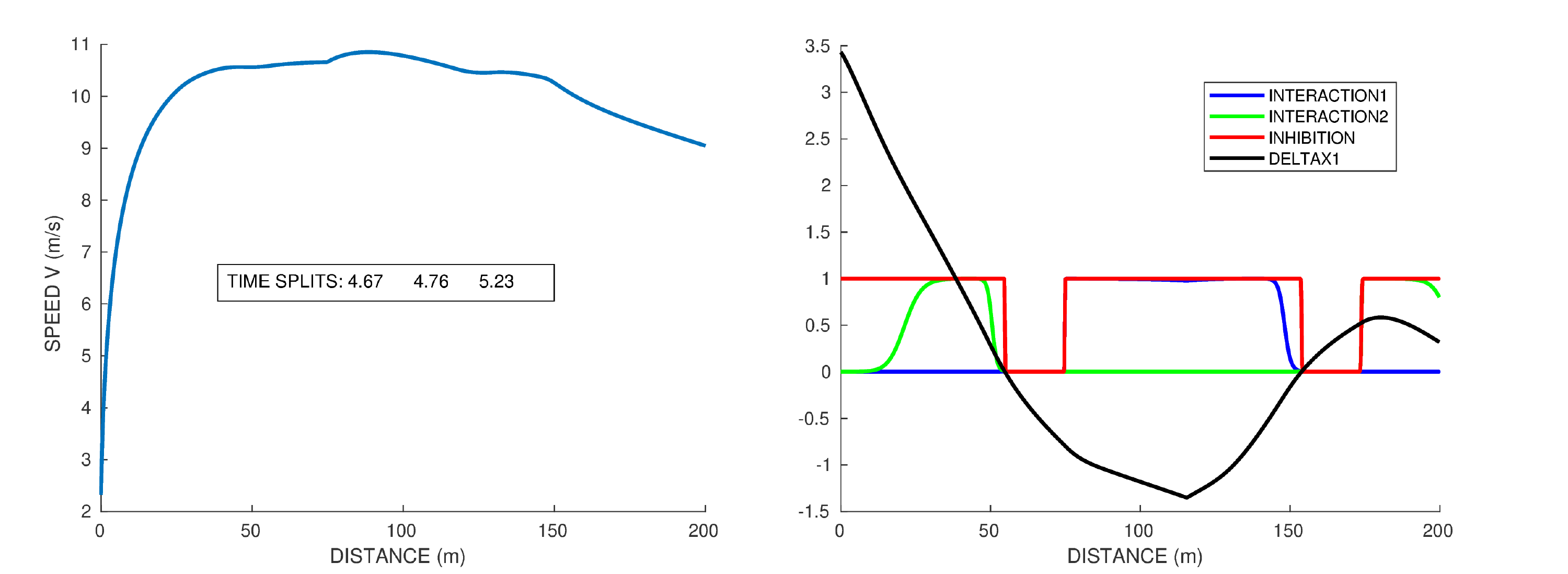}
 \end{center}
\caption{\label{fig54} Race $A_1$ vs himself at lanes 5-4, with interaction $\gamma=0.04$  and inhibition $\eta=20m$.
Left graph: speed profile and time splits of the runner at lane 5.
Right graph: distance gap and interaction term for both runners. The sign changes of the distance gap correspond to the 2 overtakings.}
\end{figure}

Then in Fig \ref{figinhib}, the interaction term is set at $\gamma=0.04$, and the inhibition frame is $\eta=20m$.
This means that the positive interaction is disabled when a runner is overtaken in the previous $20m$ of race.
Fig \ref{figinhib} shows the speed profile (left graph) and interaction / inhibition terms (right graph).
Compared to the race without inhibition in Fig \ref{fig54}, we observe a different behaviour with only one overtaking and the runner on the {\co inside} winning by $0.27s$.
Note that since we optimize the whole race, there is no reason for the race with inhibition to coincide with the race without inhibition, even before any overtaking occurs.
We observe that the inhibition (on the right graph) correctly detects the overtaking and suppresses the interaction accordingly.
This prevents the overtaken runner at lane 2 to keep up (and eventually catch up) with the one at lane 1, as we see that the distance gap increases after the overtaking.
In the race without inhibition, the overtaken runner was benefiting from the interaction right away, which allowed him to catch up and take the lead back.
With inhibition, the runner on lane 1 manages to win the race, though he is on a disadvantageous lane.
In the full race, of course, the runner in lane 2 has a neighbour on the other side which changes the total result.

There are cases where, though the inhibition $\eta=20$, there are still two overtakings.
We study for instance the 5-4 race, with the speed and force profile of the runner at lane 5 shown in Fig \ref{fig54}.
Without interaction $(\gamma=0)$, lane 5 wins without any overtakings, with final time $22.47s$.
With interaction ($\gamma=0.04$ and $\eta=20$), lane 5 still wins after 2 overtakings, with final time $22.23s$.
At the start, the runner on lane 4 benefits from the interaction due to the runner at lane 5 being ahead (staggered start).
He catches up then overtakes the {\co outer} runner, who in turn gains the interaction, catches up and overtakes the {\co inner} runner again.
At the end the {\co inner} runner, being behind, has the interaction again and is catching up with the {\co outer} runner, but too late.

We have also made simulations with runner $A_1$ vs runner $A_2$, and though runner $A_2$ is stronger in force, on some lanes,
runner $A_1$ can benefit from interaction to be able to win.

We point out that the interaction parameters can be runner dependent since some may be very sensitive to this effect and others much less.


\subsection*{Mean time per lane}
\begin{figure}[ht!]
 \begin{center}
  \includegraphics[width=0.9\textwidth]{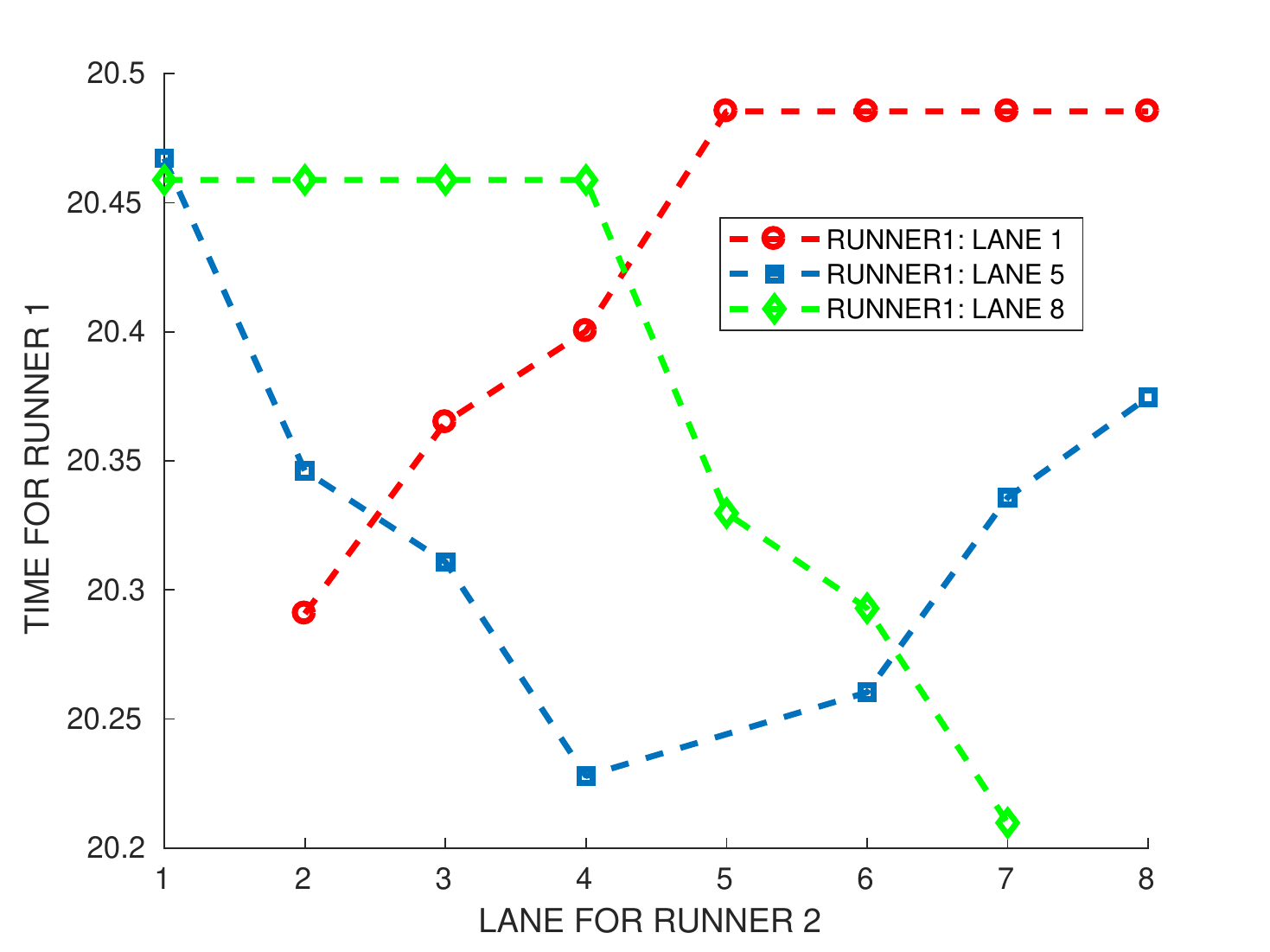}
 \end{center}
\caption{\label{figseq2}Times for athlete $A_1$ at lanes 1,5,8, running against himself.}
\end{figure}

In a real race, there are eight runners, however our model is only for two.
Therefore, we simulate a set of races with two identical runners, the first on a fixed lane, the second on each possible other lane.
We define $T_1^{k_1,k_2}$ to be the time for the winner in the race between two identical runners $A_1$ in lanes $k_1$ and $k_2$.
We want to compute the average performance at lane $i$ as the mean time
$$
\bar T_i = \frac{1}{7} \sum_{j=1..8, j\neq i} T_1^{k_1=i,k_2=j}.
$$
First, we compute the times $T_1^{i,j}$: the best times in $j$ for each $i$ are indicated in Table \ref{tab:seq2}.
In Fig \ref{figseq2}, we have plotted the times for $i=1$, 5, and 8.
The best times are obtained for the maximal interaction, namely with the second runner on an adjacent lane.
For runner $A_1$ on lane 5, his best performance is obtained with a neighbor on lane 4 rather than 6.
We recall that the model includes a lateral attenuation for the interaction, which is 0 when runners are more than 3 lanes apart.
If we compare the best time for each case, it is decreasing with the lane.

\begin{table}[ht!]
\begin{center}
\caption{\label{tab:seq2}Athlete $A_1$ at lane $k$ running against himself at lane $k-1$.
Interaction $\gamma=0.04$ with inhibition $\eta=20m$. Race time and gain with respect to solo race time.}
\begin{tabular}{l|rrrrrrr}
lane      		& 2 		& 3 		& 4 		& 5 		& 6 		& 7 		& 8\\
\hline
solo time 		&20.480 &20.475 &20.471 &20.467 &20.464 &20.461 &20.459\\
2-runner time	&20.300 &20.292 &20.283 &20.276 &20.270 &20.264 &20.259\\
time gain 		&0.180  &0.183  &0.1880 &0.191  &0.194  &0.197  &0.200
\end{tabular}
\end{center}
\end{table}

\begin{figure}[ht!]
 \begin{center}
  \includegraphics[width=0.5\textwidth]{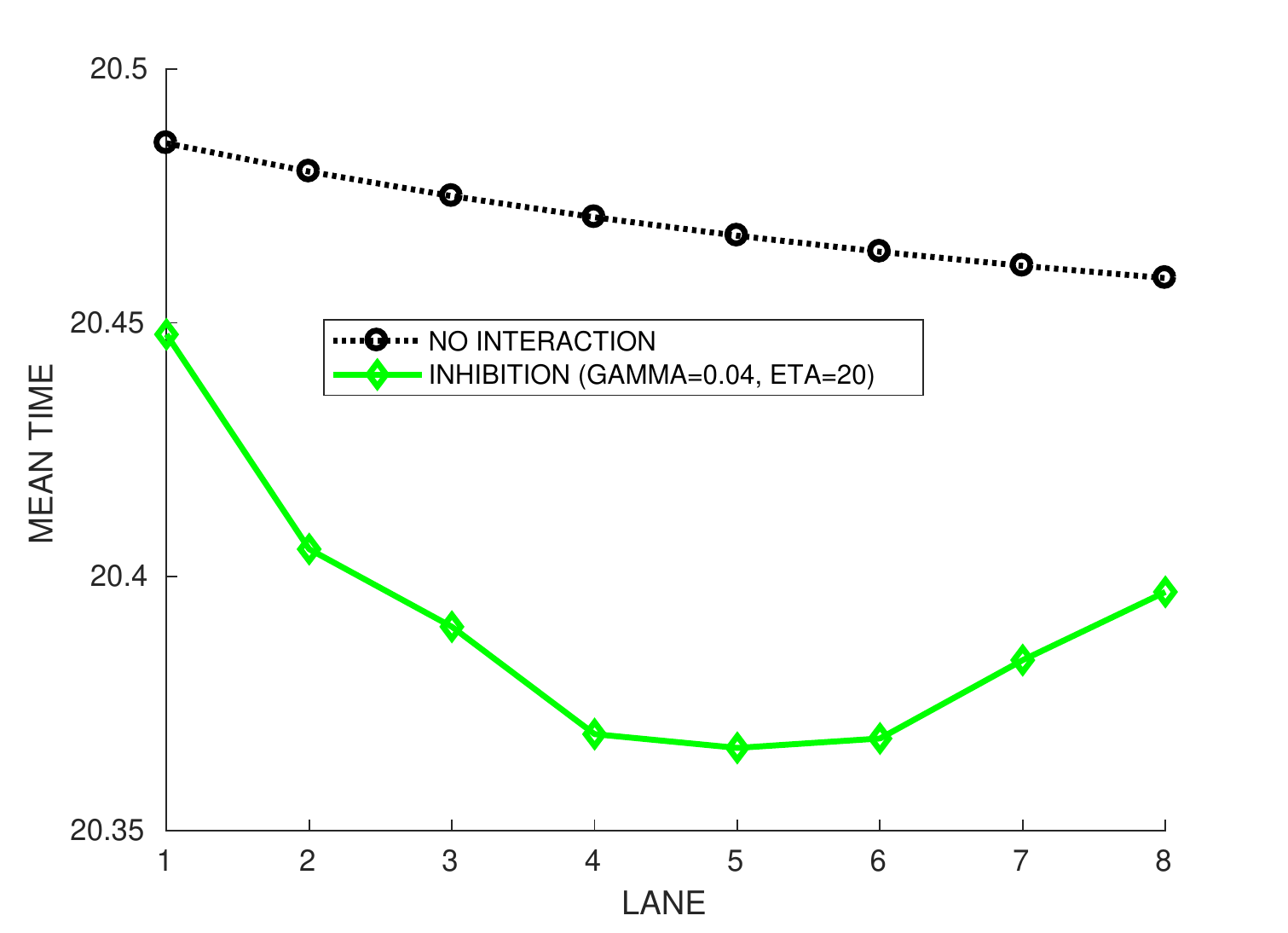}
 \end{center}
\caption{\label{figtmean} Mean times per lane for runner $A_1$ when in lane $i$ vs  himself in all other lanes.
Without interaction $(\gamma=0)$, and with interaction $\gamma=0.04$, $\eta=20$.
Lane performance (sorted by mean time): $\bar T_5 < \bar T_6 < \bar T_4 < \bar T_7 < \bar T_3 < \bar T_8 < \bar T_2 < \bar T_1$.
Gap $\bar T_1 - \bar T_8 = 0.050859$.}
\end{figure}

We show in Fig \ref{figtmean} the mean times $\bar T_i$ obtained for runner $A_1$ against himself, with an interaction weight $\gamma=0.04$ and $\eta=20$ when he runs on each lane $i$. If we look at the overall performance then lane 5 is the best, followed by lane 6, 4, 7, 3, 8, 2 and lane 1 is by far the worst. We compare with the solo case ($\gamma=0$) where of course the {\co outside} lane is the quickest.\\
The results are nicely consistent with the IAAF rules {\co for the lane drawn}. Indeed, according to the IAAF rules \cite{iaaf}, starting lanes are drawn in three lots:\\
- a first draw is made for the four best runners in the {\co center} lanes 3,4,5 and 6.\\
- a second draw is made for the next two  runners between the {\co outer} lanes 7 and 8.\\
- a last draw is made between the runners with the lowest performance to get the {\co inside} lanes 1 and 2.\\
{\co Nevertheless,} we find that the {\co inside} lanes 1,2 are a real disadvantage, the more so as if the runners are not as strong.

In \cite{ryan2003} the authors recall some average time data for Olympics 1996 and 2000, and World Championship 2001:
they indicate an average time gap of $0.16s$ between {\co inside} lanes 1 and 2 and {\co outside} lanes 7 and 8.
We obtain a smaller gap of $0.047s$, which may be due to the fact that we consider identical runners in our simulations,
while in actual races the athletes in the {\co outside} lanes were supposedly stronger than those in the {\co inside} lanes.

\section*{Conclusion}

In this paper, we have studied how the geometry of the track and the psychological interaction between runners affect performances.
We have introduced an optimal control model taking into account the centrifugal force as a limiting factor for the maximal propulsive force.
We couple this with a new model describing the positive interaction exerted by a runner close ahead and the delay to benefit
from it after being overtaken.
We carry out numerical simulations for several runner profiles on different track shapes.
The results indicate that the track with the shortest {\co straights} is the quickest for strong athletes in {\co outside} lanes.
The so-called standard track ({\co two straights  and half circles}) yields the best performances overall.
The double bend tracks with longer {\co straights} (DB2) are significantly slower.
In particular running on lane 1 on the DB2 track appears to be an overwhelming disadvantage.

Furthermore, the combination of the centrifugal and interaction effects leads to the center lanes being the most favorable,
followed by the {\co outside} lanes, with the {\co inside} lanes being the worst.
These results fit very well with { the IAAF rules for lane draws}, which follow this preference order.

\appendix

\section*{Track shape details}
\label{appendix:track_shapes}
Note that each runner is assumed to run at a distance of $30cm$ from the {\co inner} limit of the lane.
This is how the radius of the circular parts is set in order to obtain a $400m$ distance on lane 1.

\subsection*{Standard track}
The standard track is made up of a circular half-circle of length $l_c=115.61m$ followed by a {\co straight} of $84.39m$,
for a total distance of $200m$, which yields  $$R_1 = l_c/\pi = 36.80.$$
Since the runner is assumed to be $30cm$ away from the boundary of the lane, the radius of construction is $R_1-0.3$.

We denote by $R_k$ the radius for the runner on lane $k$.
Since the width of a lane is $1.22m$, the radius at lane $k$ is
$$R_k = R_1 + 1.22 (k-1).$$
Therefore, on a standard track, the radius is given by the expression
$$R_k(s)=R_k\ \forall s \in [0,l_c],\ R_k(s)= +\infty \ \forall s\geq l_c.$$

We denote by $\theta_k(s) \in[0,\pi]$ the angular position of the runner on the curved part on lane $k$, with convention $\theta_k(s)=\pi$ on the straight part.
The staggered start design ensures all lanes have the same total distance. This yields the starting angle
$$\theta^0_k = \frac{1.22(k-1)\pi}{R_k}.$$ This goes from 0 for lane 1 to $0.6 rad$ for lane 8.
On the curved part, the angular position of the runner on lane $k$ varies in $[\theta^0_k,\pi]$ according to
$$\theta_k(s) = \theta^0_k + \frac{s}{R_k}.$$



\subsection*{Double bend track}
As for the standard track, let us denote by $l_c$ the length of the curved part, and $k$ the lane number.
Let $R_o$ and $\phi_o$ be the radius and angular width of the outer (smaller) circles, and similarly $R_i,\phi_i$ for the inner circle.
Going from the starting position, we denote by $C_1$ the first circular part (`outer circle'), $C_2$ the second one (`inner circle'),
$C_3$ the second `outer circle', and $S$ the straight part.
With $\mu_k=1.22(k-1)$ the radius adjustment for each lane, the abscissa limits for $C_1$ and $C_2$ are
$$
s_1 = l_c-(R_o+\mu_k)\phi_o-(R_i+\mu_k)\phi_i,   \quad  s_2 = l_c-(R_o+\mu_k)\phi_o.
$$

Finally we have the radius expression
$$
\left \lbrace
\begin{array}{ll}
R_k(s)=R_o+\mu_k, & s \in C_1 = [0,s_1],\\
R_k(s)=R_i+\mu_k, & s \in C_2 = [s_1,s_2],\\
R_k(s)=R_o+\mu_k, & s \in C_3 = [s_2,l_c],\\
R_k(s) = +\infty, & s \in S = [l_c,d].
\end{array}
\right .
$$

Denoting by $ \theta^0_k=\frac{\mu_k \pi}{R_o} $ the starting angular position, the angle after running $s$ meters is
$$
\left \lbrace
\begin{array}{lr}
\theta_k(s)= \theta^0_k +  s/(R_o + \mu_k), & s \in C_1\\
\theta_k(s)= \phi_o + (s-s_o)/(R_i+\mu_k), & s \in C_2\\
\theta_k(s)= \pi - (l_c-s)/(R_o+\mu_k), & s \in C_3
\end{array}
\right .
$$

\subsection*{Clothoid track}
Let us study the design of a modified standard track in which the circular part of radius $R$ is bracketed by two smoother junctions of length $\bar l$ with continuous curvature.
We choose to use a junction whose curvature is linear with respect to the distance, called a clothoid (also known as Euler curve or Cornu spiral).
The angle with the tangent $\varphi$ is   $$\quad\quad \varphi(s) = \frac{s c(s)}{2}$$ where $c(s)$ is the curvature at distance $s$.
Then, since the total angle for the circular part and the two clothoid junctions is $\pi$,
and using the same notations as before that is $l_s$ is the length of the straight part and $l_c$ the length on the circle, we find
$$ \frac{l_c}{R} + 2\varphi(\bar l) = \pi.$$
Since the total angle for one clothoid is $2\varphi(\bar l) = \bar l/R$,
this equation leads to $l_c + \bar l = R \pi$. Moreover $l_s+l_c+2\bar l=d$ where $d$ is the distance of the race, that is $200m$ in our case, thus $R\pi= d - l_s -\bar l$.
We find therefore that when there is a junction with a clothoid, the radius of the circular part gets smaller than in the case of a full half circle.

On the clothoid, for $s \in [s_{begin},s_{end}]$ , the expression of the curvature is linear:
$$ c(s) = c_{begin} \frac{s_{end}-s}{s_{end}-s_{begin}} + c_{end} \frac{s-s_{begin}}{s_{end}-s_{begin}}.$$
In our case the clothoids will join the straight part (curvature $0$) and circular part (curvature $1/R$).\\

Similarly to the double bend track, we denote by $C_1,C_2,C_3,S$ respectively the first clothoid, circular, second clothoid, and straight parts.
We denote by $l_s$ the {\co straight} length, $R, l_c$ the radius and length of the circular part, and $\bar l_{1,2}$ the length of the two clothoid junctions.
As before, we call $R_k = \frac{l_c+\bar l}{\pi}+ 1.22(k-1)$ and $l_{c,k} = l_c (1 + 1.22(k-1)/R)$ the radius and length of the circular part at lane $k$.
On lane $1$, both clothoids have same length $\bar l$, while for $k>1$ the first clothoid is shorter in order to keep the same total length.
The second clothoid has full length $\bar l_{k,2} = \bar l_k = \bar l (1 + 1.22(k-1)/R)$.
Thus the first clothoid has length $\bar l_{k,1} = d/2 - l_s - l_c - \bar l_{k,2}$.\\

Taking these variable lengths into account, the curvature at lane $k$ after running $s$ meters is
$$
\left \lbrace
\begin{array}{ll}
c_k(s) = \frac{1}{R_k} \frac{s + \bar l_k-\bar l_{k,1}}{\bar l_k},                     & s \in C_1 = [0,\bar l_{k,1}],\\
c_k(s) = 1 / R_k,                                                                      & s \in C_2 = [\bar l_{k,1},\bar l_{k,1} + l_{c,k}],\\
c_k(s) = \frac{1}{R_k} \frac{\bar l_{k,1} + l_{c,k} + \bar l_{k,2} - s}{\bar l_{k,2}}, & s \in C_3 = [\bar l_{k,1} + l_{c,k},\bar l_{k,1} + l_{c,k} + \bar l_{k,2}],\\
c_k(s) = 0,                                                                            & s \in S = [\bar l_{k,1} + l_{c,k} + \bar l_{k,2},d].
\end{array}
\right .
$$

\subsection*{Distance gap between two runners}
\label{appendix:distance_gap}
For the interaction term, we need to define the relative distance between the runner on lane 1 and on lane 2, taken by convention at time $y_1(s)$:
$$\rho(s) = x_1(y_1(s))-x_2(y_1(s)) = s - x_2(y_1(s)).$$
Thus, runner 1 is ahead of runner 2 at time $y_1(s)$ when $\rho(s) > 0$, and behind otherwise.
However the term $x_2(y_1(s)$ (of derivative $\dot{x}_2(y_1(s))=1/\dot{y}_2 (y_2^{-1}(y_1(s))$) is rather difficult to handle numerically.
Therefore, we replace $\rho(s)$ with a more handy approximation of the distance between runners, namely the mean velocity multiplied by the times difference
$$
r(s) = (y_2(s)-y_1(s)) \frac{v_1(s)+v_2(s)}{2}.
$$

On a curved track, this approximation is adjusted by projecting the two runners on a median circle,
while also taking into account the staggered start on different lanes:
$$
r(s) = (y_2(s)-y_1(s)) \frac{v_1(s)+v_2(s)}{2} + (\theta_1(s)-\theta_2(s)) \frac{2}{c_1(s)+c_2(s)},
$$
where $c_i(s)=1/R_i(s)$ is the curvature at distance $s$.


\section*{Acknowledgments}
The authors would like to thank Cecile Appert and Henk Hillhorst for introducing the delay problem for the inhibition of runners interaction.
They also acknowledge enlightening discussions on the topic with Thorsten Emig, Christine Hanon, Jean-Pierre Nadal and Guillaume Rao.
They are very grateful to Aurelien Alvarez for providing details on clothoids.


\end{document}